\documentclass[12pt]{article}
\usepackage{amsmath}
\usepackage{amsthm}
\usepackage{wasysym}
\usepackage{graphicx}
\usepackage{color}%
\usepackage[doublespacing]{setspace}
\usepackage{lipsum}
\usepackage[lmargin=3cm,rmargin=3cm,tmargin=3cm,bmargin=3cm]{geometry}
\usepackage{multirow}
\usepackage{hyperref}
\usepackage{mathtools}
\usepackage{breqn}
\usepackage{threeparttable}
\usepackage[utf8]{inputenc}
\usepackage{tocloft}
\usepackage{cite}
\usepackage{enumitem}

\begin{document}

\begin{center}
{\rm \bf \Large{Revisiting the Persson theory of elastoplastic contact: A simpler closed-form solution and a rigorous proof of boundary conditions}}
\end{center}

\begin{center}
{\bf Yang Xu$^{a}$\footnote{Corresponding author: Yang.Xu@hfut.edu.cn}, Xiaobao Li$^b$, Daniel M. Mulvihill$^c$}\\
{
$^a$School of Mechanical Engineering, Hefei University of Technology, Hefei, 230009, China\\
$^b$School of Civil Engineering, Hefei University of Technology, Hefei, 230009, China\\
$^c$Materials and Manufacturing Research Group, James Watt School of Engineering, University of Glasgow, Glasgow, G12 8QQ, UK
}
\end{center}

\begin{center}
{\bf Abstract}
\end{center}

Persson's theory of contact is extensively used in the study of the purely normal interaction between a nominally flat rough surface and a rigid flat. In the literature, Persson's theory was successfully applied to the elastoplastic contact problem with a scale-independent hardness $H$. However, it yields a closed-form solution, $P(p, \xi)$, in terms of an infinite sum of sines. In this study, $P(p, \xi)$ is found to have a simpler form which is a superposition of three Gaussian functions. A rigorous proof of the boundary condition $P(p=0, \xi)=P(p=H, \xi) = 0$ is given based on the new solution.

\begin{flushleft}
\end{flushleft}

\section{Introduction}
In the study of the purely normal contact between a nominally flat rough surface and a rigid flat, there are two major theoretical approaches, namely, the multi-asperity contact model (e.g., the Greenwood and Williamson (GW) model\cite{Greenwood66}) and Persson's theory of contact \cite{Persson01}. In the GW model, as well as other related multi-asperity contact models, the physical basis for the rough surface contact is clear, i.e., solid interaction occurs at the summits of higher asperities. Therefore, multi-asperity contact models rely on the analytical asperity models (e.g., Hertzian contact theory \cite{Johnson87}) or empirical models (e.g., Jackson and Green model \cite{Jackson05}). One major assumption adopted in the multi-asperity contact models is that each contacting asperity can be studied individually without considering its interaction with nearby asperities. However, this assumption restricts the application of the multi-asperity contact models in the low load range where asperity interaction and coalescence are largely avoided. So far, several attempts have been made to overcome this limitation \cite{Ciavarella06, Afferente12, Xu14, Xu17}.

A major new development in this field arrived some two decades ago with the publication of Persson's theory of contact\cite{Persson01}. Since then, it has gradually become one of the dominant theoretical approaches in the area of rough surface contact. Here, unlike multi-asperity contact models, the rough surface contact problem is solved by Persson's theory in the probability domain. Assuming the contact pressure, $p$, within the real area of contact is a stochastic variable, Persson derived a diffusion equation of a scale-dependent probability density function (PDF) $P(p, \xi)$ where the scale $\xi = \kappa_s/\kappa_l$ is the ratio of the upper cut-off frequency $\kappa_s$ to the lower cut-off frequency $\kappa_l$ associated with the power spectrum density of the rough surface topography. As the scale $\xi$ increases, higher frequency components are added to the roughness topography. As a matter of fact, the corresponding probability density function $P(p, \xi)$ is broadened and evolves following the diffusion equation \cite{Persson01}. Compared with the multi-asperity contact models, the asperity interaction is effectively included in Persson's theory of contact. Moreover, the theory covers the entire load range of rough surface contact from the first touch to nearly complete contact.

Persson's theory of contact was first applied by Persson \cite{Persson01, Persson01a} to the elastoplastic contact problem. To account for yielding, the contact pressure is truncated at $p = H$ where $H$ is a scale-independent hardness. Therefore, the elastoplastic material model used in Persson’s theory \cite{Persson01, Persson01a} is indeed a constant hardness model. It should not be confused with the elastoplastic models (e.g., elastic-perfectly plastic model) used in the solid mechanics community. The probability of the elastic portion of the contact pressure is assumed to satisfy the diffusion equation. Persson's theory of contact yields an analytical solution of $P(p, \xi)$ in a Fourier series (in terms of an infinite sum of sines). An attempt to include the size-dependent plasticity has been made by Persson \cite{Persson06} using a scale-dependent hardness $H(\xi)$. Venugopalan \emph{et al.} \cite{Venugopalan19} found good agreement between Persson's theory and discrete dislocation plasticity modelling. The agreement deteriorates as root mean square roughness increases. This disagreement probably arises for two possible reasons \cite{Venugopalan19}: (1) Persson's theory does not account for the strain hardening, and (2) an inaccurate estimation of the hardness using $H = 3 \sigma_Y$ where $\sigma_Y$ is the initial yield strength.

In the present study, Persson's theory of elastoplastic contact with a scale-independent hardness is revisited. Inspired by the mirror Gaussian solution \cite{Manners06, Wang17} of the linear elastic contact problem, a simpler analytical solution of $P(p, \xi)$ is found which is a superposition of three Gaussian functions.

\section{Elastoplastic Contact}\label{sec:Elastoplastic}
\begin{figure}[h!]
  \centering
  \includegraphics[scale=0.6]{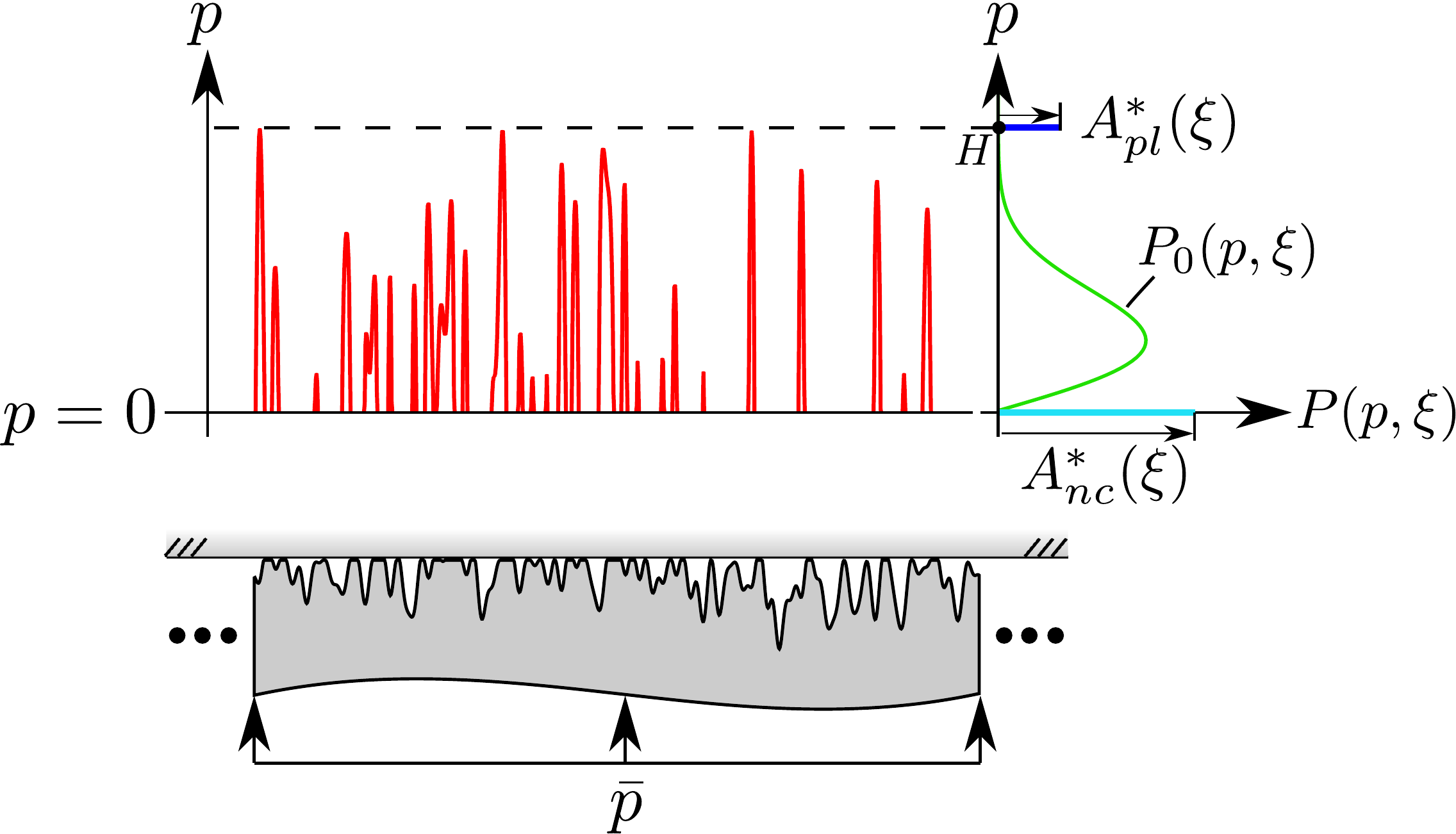}
  \caption{Schematic of the deformed interface, the elastoplastic contact pressure and the corresponding PDF $P(p, \xi)$.}\label{fig:Fig_1}
\end{figure}

Consider a purely normal contact between a nominally flat rough surface and a rigid flat, see Fig. \ref{fig:Fig_1}. The contact pair is subjected to a uniform normal traction $\bar{p}$ applied remotely. The scale-dependent PDF of the contact pressure, $P(p, \xi)$, is discontinuous at $p=0$ and $p = H$, i.e., $P(p = 0, \xi) \neq P(p \to 0^+, \xi)$ and $P(p = H, \xi) \neq P(p \to H^-, \xi)$. This is because the non-contact area and plastically deformed area is associated with $p = 0$ and $p = H$, respectively. Inspired by a similar form proposed by Persson \cite{Persson18} for an adhesive contact problem, a discontinuous PDF $P(p, \xi)$ can be formulated as follows
\begin{equation}\label{E:PDF_partial}
  P(p, \xi) = \delta(p) A_{nc}^*(\xi) + \delta(p - H) A_{pl}^*(\xi) + \left[ H(p) - H(p - H) \right] P_0(p, \xi),
\end{equation}
where $\delta(p)$ and $H(p)$ are \emph{Dirac} and \emph{Heaviside} functions, respectively. The first two terms in the right hand side of Eq. \eqref{E:PDF_partial} represent the PDF of contact pressure $p = 0$ and $p = H$, respectively. The last term is associated with the PDF of the elastic contact pressure $ p \in (0, H)$ which is denoted by $P_0(p, \xi)$. Outside $p \in (0, H)$, $P_0(p, \xi) = 0$. $A_{nc}^*(\xi)$, $A_{el}^*(\xi) = \int_{0^+}^{H^-} P_0(p, \xi) dp$ and $A_{pl}^*(\xi)$ are the ratios of the non-contact, elastic and plastic contact areas to the nominal contact area, respectively. The PDF $P(p, V)$ is automatically vanishing outside the nominal range $p \in [0, H]$. A schematic plot of $P(p, V)$ can be found in Fig. \ref{fig:Fig_1}.

Based on Persson's theory of elastoplastic contact \cite{Persson01, Persson01a}, the PDF of elastic contact pressure $P_0(p, \xi)$ satisfies the following diffusion equation:
\begin{equation}\label{E:SDE_partial_1}
\frac{\partial}{\partial \xi} P_0(p, \xi) = \frac{1}{2} \frac{d V}{d \xi} \frac{\partial^2}{\partial p^2} P_0(p, \xi) ~~~ p \in (0, H) ~~~\xi \geq 1,
\end{equation}
where $V = \langle p_c^2 \rangle$ is the variance of the elastic contact pressure $p_c$ at the stage of complete contact \cite{Manners06}. For a self-affine isotropic roughness with a deterministic power spectrum density \cite{Persson05, Yastrebov15}, $V(\xi)$ monotonically increases with the scale $\xi$. Neglecting the finite size effect (i.e., $\kappa_l \to 0$), roughness vanishes so that $V(\xi = 1) = 0$.

Assuming $V(\xi)$ monotonically increases with $\xi$ with $V(\xi = 1) = 0$, then $V(\xi)$ is a one to one mapping from $\xi$ to $V$, and we can replace $\xi$ with $V$ in Eq. \eqref{E:SDE_partial_1}, i.e., $P(p, \xi) = P(p, V)$. The corresponding diffusion equation is exactly what Manners and Greenwood derived \cite{Manners06}
\begin{equation}\label{E:SDE_partial}
\frac{\partial}{\partial V} P_0(p, V) = \frac{1}{2} \frac{\partial^2}{\partial p^2} P_0(p, V) ~~~ p \in (0, H) ~~~V \geq 0.
\end{equation}
Since the finite size effect is neglected, we can expect that the contact surface becomes perfectly flat with an infinite size as $V(\xi = 1) = 0$ and solid contact occurs over the entire nominal contact area with a uniform pressure of magnitude $\bar{p}$. Therefore, the initial condition of Eq. \eqref{E:SDE_partial} is simply a Dirac function \cite{Persson01a}
\begin{equation}\label{E:Initial_condition}
P_0(p, V = 0) = \delta(p - \bar{p}).
\end{equation}
The corresponding boundary conditions at $p = 0^+$ and $p = H^-$ was given by Persson with a less rigorous proof \cite{Persson01a, Persson06}:
\begin{equation}\label{E:BCs_elastoplastic}
P_0(p = 0^+, V) = P_0(p = H^-, V) = 0.
\end{equation}
Eqs. (\ref{E:SDE_partial}-\ref{E:BCs_elastoplastic}) are solved below by two approaches: the former one is the Fourier series method proposed by Persson \cite{Persson01a}, and the latter one is a new method we propose here based on the superposition of three Gaussian distributions. A rigorous proof of the boundary conditions in Eq. \eqref{E:BCs_elastoplastic} is given in Section \ref{sec:Discussion}.

\subsection{Fourier series method}\label{subsec:FSM}
A general solution of Eq. \eqref{E:SDE_partial} maybe given in Fourier series form as follows:
\begin{equation}\label{E:P0_FS}
P_0(p, V) = \sum_{n=1}^{\infty} B_n(V) \sin \left(\frac{n \pi p}{H}\right),
\end{equation}
which automatically satisfies the boundary conditions in Eq. \eqref{E:BCs_elastoplastic}. Substituting Eq. \eqref{E:P0_FS} into Eqs. \eqref{E:SDE_partial} and \eqref{E:Initial_condition}, we can have an explicit form of $P_0(p, V)$ in terms of an infinite sum of sines\cite{Persson01, Persson01a}:
\begin{equation}\label{E:P0_FS_1}
P_0(p, V) = \frac{2}{H} \sum_{n=1}^{\infty} \sin\left( \frac{n \pi \bar{p}}{H} \right) \sin\left(\frac{n \pi p}{H}\right) \exp\left[-\frac{1}{2} \left( \frac{n \pi}{H} \right)^2 V \right].
\end{equation}

\subsection{Superposition method}\label{subsec:MG}
\begin{figure}[h!]
  \centering
  \includegraphics[scale=0.6]{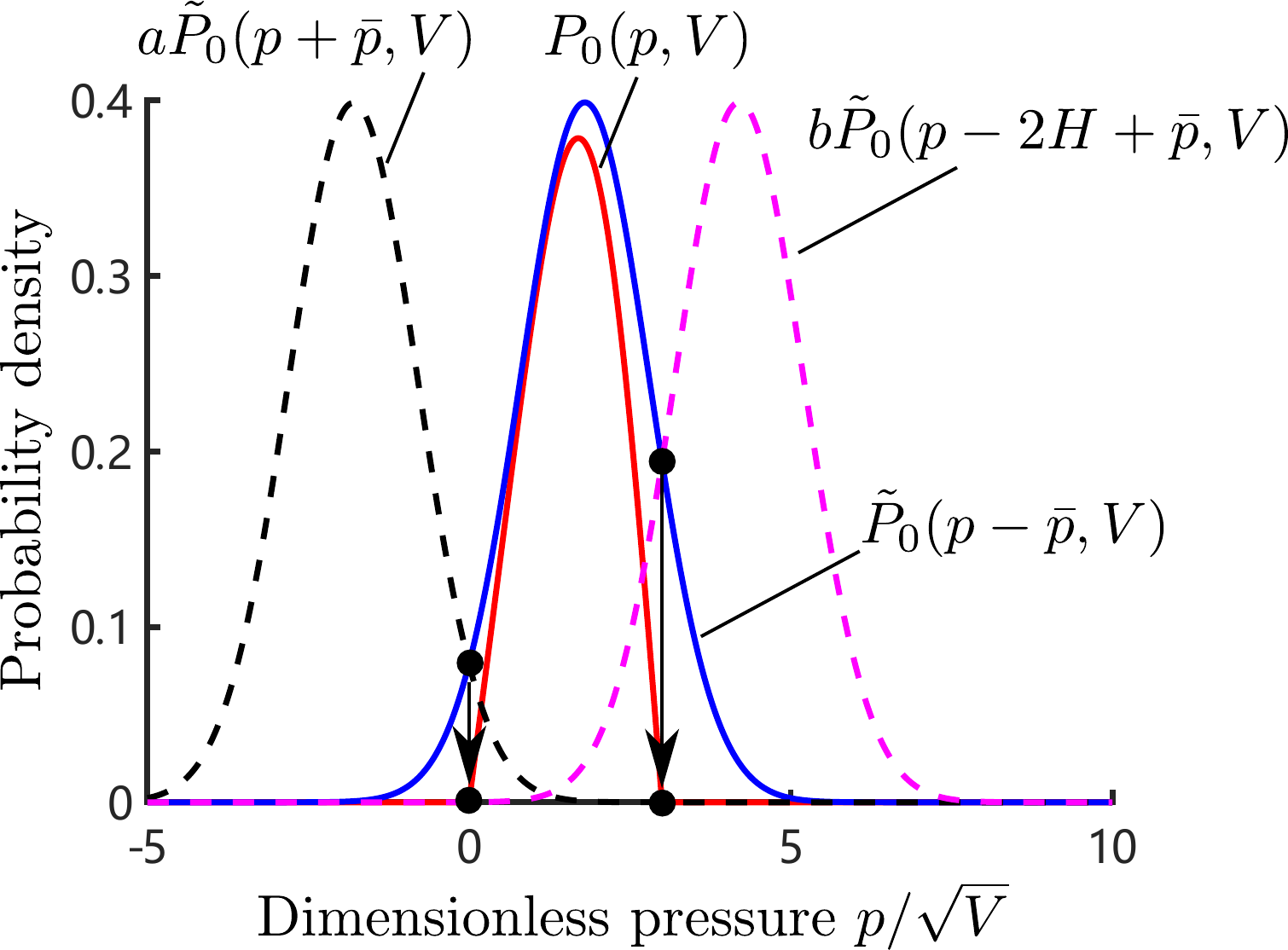}
  \caption{Schematic illustration of $P_0(p, V)$ and its three Gaussian functions where $\bar{p}/\sqrt{V} = 1.8$, $H/\sqrt{V} = 3$ and $V = 1$}\label{fig:Fig_2}
\end{figure}

The superposition method relies on a specific solution $\tilde{P}_0(p, V)$ of the diffusion equation \cite{Manners06}
\begin{equation}\label{E:Gaussian_distribution}
  \tilde{P}_0(p, V) = \frac{1}{\sqrt{2 \pi V}} \exp \left(- \frac{p^2}{2 V}\right),
\end{equation}
which is a Gaussian function. $P_0(p, V)$ can be built using a superposition of three $\tilde{P}_0(p, V)$ with two unknown coefficients:
\begin{equation}\label{E:P0_guess_elastoplastic}
  P_0(p, V) = \tilde{P}_0(p - \bar{p}, V) - a \tilde{P}_0(p + \bar{p}, V) - b \tilde{P}_0(p - 2H + \bar{p}, V).
\end{equation}
This idea is inspired by the mirror Guassian solution of the linear elastic contact \cite{Manners06, Wang17}. It is obvious that the diffusion equation in Eq. \eqref{E:SDE_partial} is automatically satisfied by the solution in Eq. \eqref{E:P0_guess_elastoplastic}. The boundary conditions in Eq. \eqref{E:BCs_elastoplastic} is enforced in this solution using two constants $a$ and $b$. An illustration of the superposition method can be found in Fig. \ref{fig:Fig_2}.

Asymptotic solutions of $a$ and $b$ can be found when $\bar{p} \to 0^+$ and $\bar{p} \to H^-$. As $\bar{p} \to 0$, $P_0(p, V) = (1 - a) \tilde{P}_0(p, V) - b \tilde{P}_0(p-2H, V) = 0$, and we can get $a = 1$ and $b = 0$. Similarly, as $\bar{p} \to H$, $P_0(p, V) = (1 - b) \tilde{P}_0(p - H, V) - a \tilde{P}_0(p + H, V) = 0$, and we have $a = 0$ and $b = 1$. Notice that this asymptotic analysis is independent of the boundary conditions.

As $V \to 0^+$, Eq. \eqref{E:P0_guess_elastoplastic} deduces to
\[
  P_0(p, V \to 0^+) = \delta(p - \bar{p}) - \delta(p + \bar{p}) - \delta(p - 2H + \bar{p}).
\]
Since $\bar{p}, p \in [0, H)$, $\delta(p + \bar{p})$ and $\delta(p - 2H + \bar{p})$ are zero. Therefore, the initial condition is satisfied.

\begin{figure}[h!]
\centering
\includegraphics[scale=0.7]{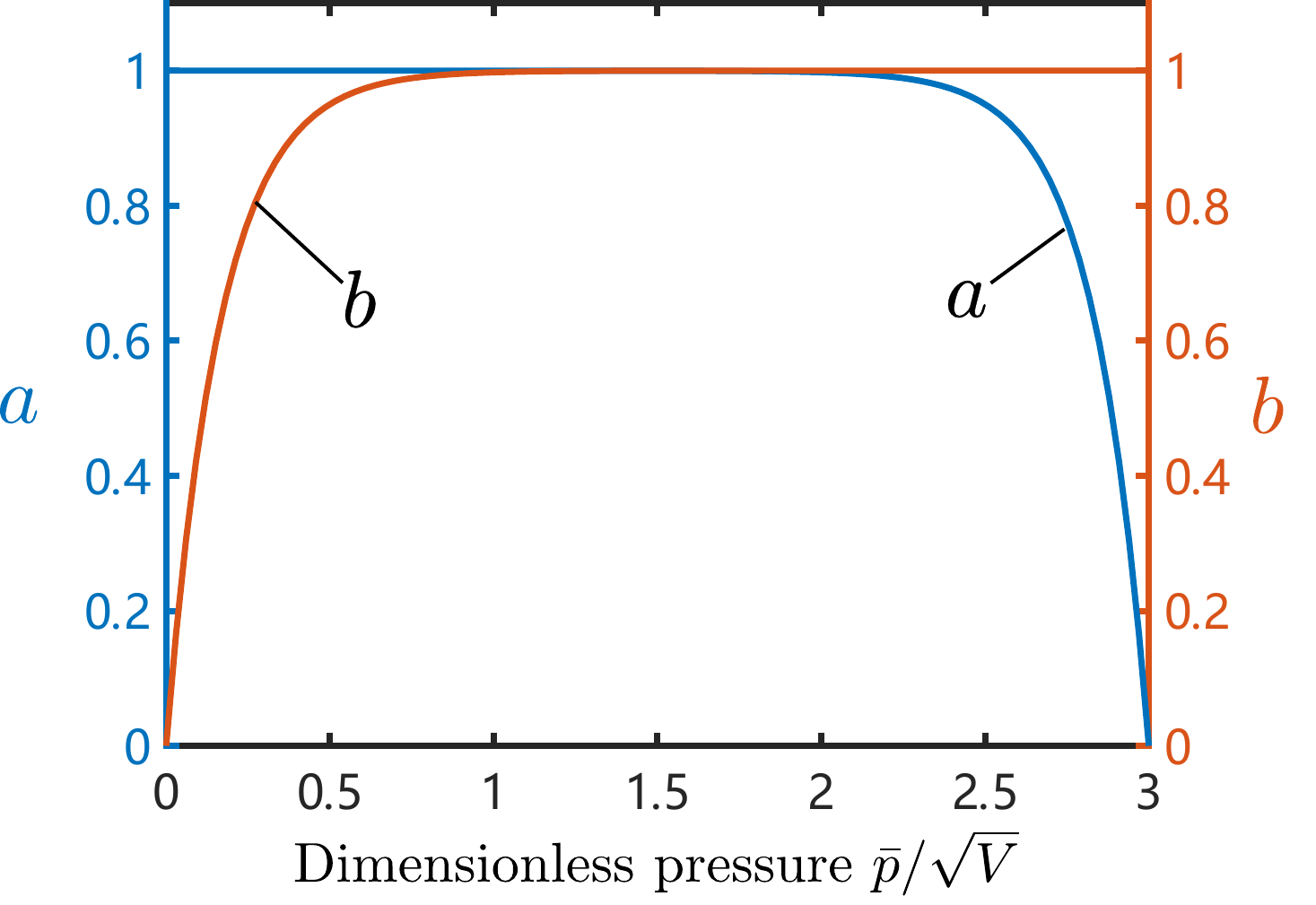}
\caption{Numerical solutions of $a$ and $b$ where $H/\sqrt{V} = 3$}\label{fig:Fig_3}
\end{figure}

Enforcing boundary conditions in Eq. \eqref{E:BCs_elastoplastic} results in two linear equations
\begin{align}
(1 - a)\tilde{P}_0(\bar{p}, V) - b \tilde{P}_0(-2H + \bar{p}, V) &= 0, \label{E:Nonlinear_elastoplastic_1} \\
(1 - b)\tilde{P}_0(H - \bar{p}, V) - a \tilde{P}_0(H + \bar{p}, V) &= 0. \label{E:Nonlinear_elastoplastic_2}
\end{align}
After solving the above two linear equations, explicit forms of $a$ and $b$ are derived
\begin{align}
a = & \frac{\tilde{P}_0(H - \bar{p}, V) \left[ \tilde{P}_0(\bar{p}, V) - \tilde{P}_0(2H - \bar{p}, V)\right]}{\tilde{P}_0(H - \bar{p}, V)\tilde{P}_0(\bar{p}, V) - \tilde{P}_0(H + \bar{p}, V)\tilde{P}_0(2H - \bar{p}, V)}, \label{eq:a_solu} \\
b = & \frac{\tilde{P}_0(\bar{p}, V) \left[ \tilde{P}_0(H - \bar{p}, V) - \tilde{P}_0(H + \bar{p}, V)\right]}{\tilde{P}_0(H - \bar{p}, V)\tilde{P}_0(\bar{p}, V) - \tilde{P}_0(H + \bar{p}, V)\tilde{P}_0(2H - \bar{p}, V)}. \label{eq:b_solu}
\end{align}
$a$ and $b$ monotonically decreases and increases, respectively, with $\bar{p}/\sqrt{V}$ for a given hardness $H$, see Fig. \ref{fig:Fig_3}.

The final form of $P_0(p, V)$ is given below
\begin{equation}\label{E:final_P0_elastoplastic}
P_0(p, V) = \frac{1}{\sqrt{2 \pi V}} \left\{ \exp \left[ -\frac{(p - \bar{p})^2}{2 V}\right] - a \exp \left[ -\frac{(p + \bar{p})^2}{2 V}\right] - b \exp \left[ -\frac{(p - 2H + \bar{p})^2}{2 V}\right]\right\},
\end{equation}
where $a$ and $b$ are given in Eqs. \eqref{eq:a_solu} and \eqref{eq:b_solu}, respectively. As $H \to \infty$, the third Gaussian term in Eq. \eqref{E:final_P0_elastoplastic} vanishes, and Eq. \eqref{eq:a_solu} results in $a = 1$. Therefore, Eq. \eqref{E:final_P0_elastoplastic} deduces to a mirror Gaussian solution of linear elastic contact \cite{Persson06, Manners06}
\begin{equation}\label{E:final_P0_elastic}
P_0(p, V) = \frac{1}{\sqrt{2 \pi V}} \left\{ \exp \left[ -\frac{(p - \bar{p})^2}{2 V}\right] - \exp \left[ -\frac{(p + \bar{p})^2}{2 V}\right] \right\}.
\end{equation}
Note that it is not straightforward to deduce the elastic solution when $P_0(p, V)$ is in the form of Fourier series, see Appendix B in \cite{Persson01}.

\subsection{$A_{el}^*$, $A_{pl}^*$ and $A_{nc}^*$}
Normalization of $P(p, V)$ results in the following equality
\begin{equation}\label{E:Normalization_Elastoplastic}
  A_{nc}^*(V) + A_{el}^*(V) + A_{pl}^*(V) = 1.
\end{equation}
Integrating both sides of Eq. \eqref{E:SDE_partial} over $p \in [0, H]$ and resorting to Eq. \eqref{E:Normalization_Elastoplastic}, we can have explicit forms of $A_{nc}^*$ and $A_{pl}^*$ \cite{Persson01a}:
\begin{align}
  A_{nc}^*(V) & = ~~\frac{1}{2} \int_0^{V} \left[ \frac{\partial}{\partial p} P_0(p, V') \right]\bigg|_{p = 0} dV', \label{E:Anc_elastoplatic} \\
  A_{pl}^*(V) & = -\frac{1}{2} \int_0^{V} \left[ \frac{\partial}{\partial p} P_0(p, V') \right]\bigg|_{p = H} dV'. \label{E:Apl_elastoplatic}
\end{align}
Substituting Eq. \eqref{E:P0_FS_1} into Eqs. (\ref{E:Normalization_Elastoplastic}-\ref{E:Apl_elastoplatic}), the Fourier series method results in the following closed-form ratios \cite{Persson01a}
\begin{align}
A_{el}^*(V) &= \sum_{n = 1}^{\infty} \left[ 1 - (-1)^n \right] \frac{2}{n \pi} \sin \left( \frac{n \pi \bar{p}}{H} \right) \exp \left[-\frac{1}{2} \left( \frac{n \pi}{H} \right)^2 V \right], \label{E:final_Ael_Fourier}\\
A_{pl}^*(V) &= \sum_{n=1}^{\infty} (-1)^{n+1} \frac{2}{n \pi} \sin \left( \frac{n \pi \bar{p}}{H}\right) \left\{ 1 - \exp\left[ -\frac{1}{2} \left( \frac{n \pi}{H} \right)^2 V \right] \right\}, \label{E:final_Apl_Fourier} \\
A_{nc}^*(V) &= \sum_{n=1}^{\infty} \frac{2}{n \pi} \sin\left( \frac{n \pi \bar{p}}{H} \right) \left\{ 1 - \exp \left[-\frac{1}{2} \left( \frac{n \pi}{H} \right)^2 V \right] \right\}. \label{E:final_Anc_Fourier}
\end{align}
The superposition method gives a simpler forms as follows
\begin{align}
A_{el}^*(V) &= \frac{1 + b}{2}\text{erf}\left( \frac{H - \bar{p}}{\sqrt{2 V}} \right) + \frac{1 + a}{2} \text{erf} \left( \frac{\bar{p}}{\sqrt{2 V}} \right) - \frac{a}{2}\text{erf}\left( \frac{H + \bar{p}}{\sqrt{2 V}} \right) + \frac{b}{2}\text{erf}\left( \frac{\bar{p} - 2H}{\sqrt{2 V}} \right), \label{E:final_Ael_elastoplastic} \\
A_{pl}^*(V) &= -\frac{1 + b}{2}\text{erf}\left( \frac{H - \bar{p}}{\sqrt{2 V}} \right) + \frac{a}{2} \text{erf} \left( \frac{H + \bar{p}}{\sqrt{2 V}} \right) + \frac{1}{2}(1 - a + b), \label{E:final_Apl_elastoplastic} \\
A_{nc}^*(V) &= -\frac{(1 + a)}{2}\text{erf}\left( \frac{\bar{p}}{\sqrt{2 V}} \right) - \frac{b}{2} \text{erf} \left( \frac{\bar{p} - 2H}{\sqrt{2 V}} \right) + \frac{1}{2}(1 + a - b). \label{E:final_Anc_elastoplastic}
\end{align}
It is easy to check that Eq. \eqref{E:Normalization_Elastoplastic} is satisfied by Eqs. (\ref{E:final_Ael_elastoplastic}--\ref{E:final_Anc_elastoplastic}). The satisfaction of Eq. \eqref{E:Normalization_Elastoplastic} by the Fourier series solutions, Eqs. (\ref{E:final_Ael_Fourier}--\ref{E:final_Anc_Fourier}), is also not straightforward. As $H \to \infty$, $a = 1$, $b = 0$, and the corresponding real contact area can be obtained directly from Eq. \eqref{E:final_Ael_elastoplastic}
\begin{equation}
A_{el}^*(V) = \text{erf} \left( \frac{\bar{p}}{\sqrt{2V}} \right).
\end{equation}

\section{Results and discussion}\label{sec:Discussion}
\begin{figure}[h!]
  \centering
  \includegraphics[scale=0.55]{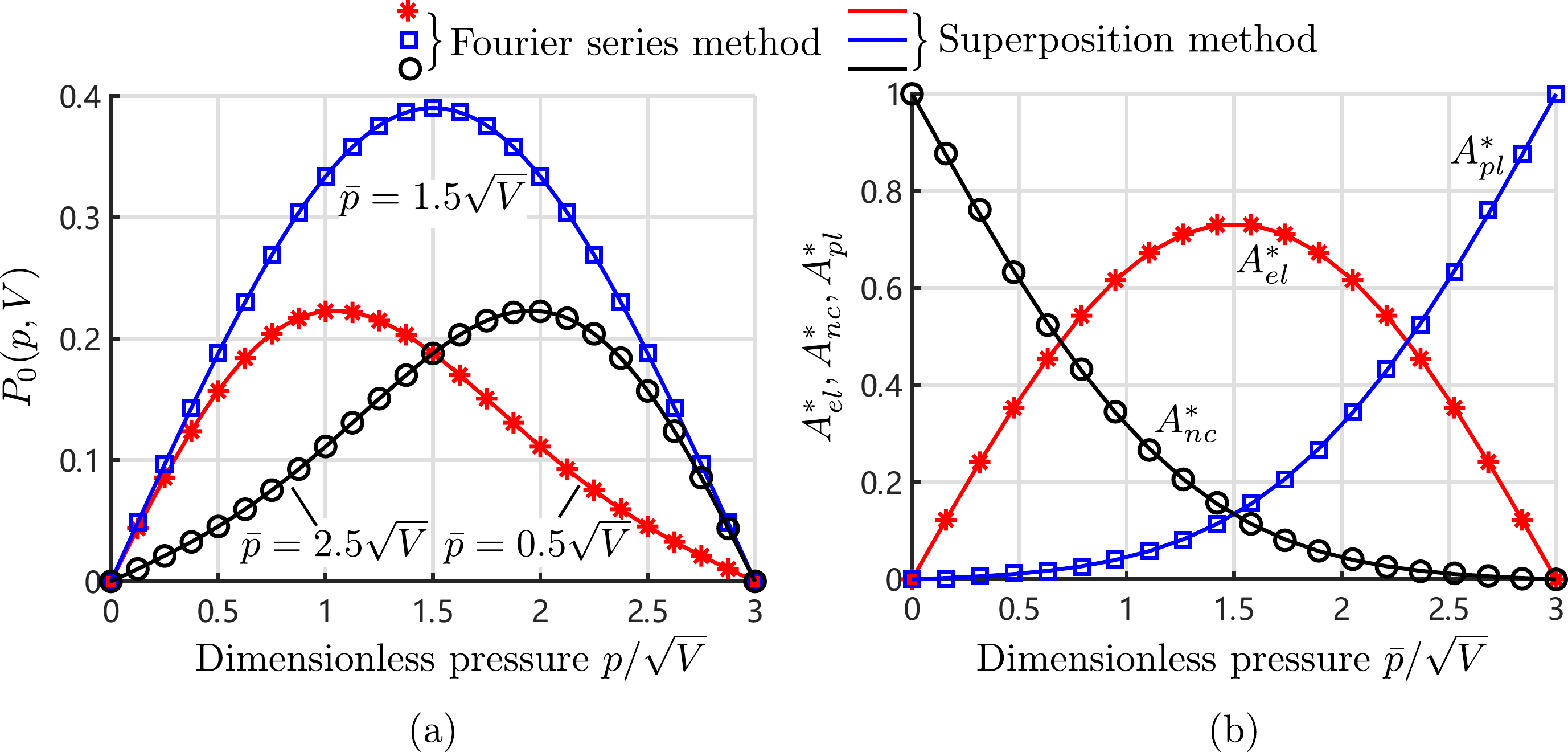}
  \caption{(a) PDF of elastic contact pressure, $P_0(p, V)$ where $\bar{p}/\sqrt{V} = 0.5, 1.5$, and $2.5$, $H/\sqrt{V} = 3$; (b) Variation of $A_{el}^*$, $A_{pl}^*$ and $A_{nc}^*$ with $\bar{p}/\sqrt{V}$ where $H/\sqrt{V} = 3$}\label{fig:Fig_4}
\end{figure}
In this section, we first compare the results of the Fourier series method with that of the superposition method. Consider the case where $\bar{p}/\sqrt{V} = 0.5, 1.5$, and $2.5$, $H/\sqrt{V} = 3$. The corresponding $P_0(p, V)$ are shown in Fig. \ref{fig:Fig_4}(a). Excellent agreement can be found between $P_0(p, V)$ predicted by Eq. \eqref{E:P0_FS_1} (Fourier series method) and Eq. \eqref{E:final_P0_elastoplastic} (superposition method). Only $10$ sine terms are used in the Fourier series method in Fig. \ref{fig:Fig_4}(a). Similar agreement can also be observed in Fig. \ref{fig:Fig_4}(b) for the variation of $A_{el}^*$, $A_{pl}^*$ and $A_{nc}^*$ with $\bar{p}/\sqrt{V}$ predicted by the Fourier series method, Eqs. (\ref{E:final_Ael_Fourier}--\ref{E:final_Anc_Fourier}), and the superposition method, Eqs. (\ref{E:final_Ael_elastoplastic}--\ref{E:final_Anc_elastoplastic}). An insufficient number of sine terms (say $10$) would result in the Fourier series solutions, $A_{pl}^*$ and $A_{nc}^*$, oscillating about the corresponding solutions of the superposition method. A larger amount of sine terms (say $1000$ in Fig. \ref{fig:Fig_4}(b)) are needed. Fig. \ref{fig:Fig_4} also shows that $P_0(p, V)|_{\bar{p} = p_0}$ is exactly the reflection of $P_0(p, V)|_{\bar{p} = H - p_0}$ about the axis $p = H/2$ where $p_0 \in (0, H)$.

In Section \ref{sec:Elastoplastic}, the boundary conditions in Eq. \eqref{E:BCs_elastoplastic} was given without a rigorous proof. In this discussion, a rigorous proof of the boundary conditions, Eq. \eqref{E:BCs_elastoplastic}, will be given with the aid of the new superposition solution given in Eq. \eqref{E:P0_guess_elastoplastic}.

The load equilibrium of elastoplastic contact can be formulated as
\begin{equation}\label{E:Load_equilibrium_elastoplastic}
  \int_{0^+}^{H^-} p P_0(p, V) dp + H A_{pl}^*(V) = \bar{p}.
\end{equation}
Since $H$ and $\bar{p}$ are scale-independent, we can get the following equation by differentiating both sides of Eq. \eqref{E:Load_equilibrium_elastoplastic} with respect to $V$:
\begin{equation}\label{E:Load_equilibrium_elastoplastic_1}
  \int_{0}^{H} p \frac{\partial}{\partial V} P_0(p, V) dp + H \frac{d}{d V} A_{pl}^*(V) = 0.
\end{equation}
Using Eq. \eqref{E:SDE_partial} and \eqref{E:Apl_elastoplatic}, we can replace $\displaystyle \frac{\partial}{\partial V} P_0(p, V)$ and $\displaystyle \frac{d}{d V} A_{pl}^*(V)$ with $\displaystyle \frac{1}{2} \frac{\partial^2}{\partial p^2} P_0(p, V)$ and $\displaystyle -\frac{1}{2} \left[ \frac{\partial}{\partial p} P_0(p, V)  \right]\bigg|_{p = H} $, respectively. Using integration by parts, Eq. \eqref{E:Load_equilibrium_elastoplastic_1} eventually becomes
\begin{equation}\label{E:BCs_elastoplastic_new}
  P_0(p=0^+, V) = P_0(p=H^-, V) = C(V),
\end{equation}
where $C$ is a function of $V$ only.

Persson \cite{Persson06} argued that $P_0(p=0^+, V) = 0$ which proved in linear elastic contact problem is still held in the elastoplastic problem. Thus, $C(V) = 0$ and eventually, boundary condition in Eq. \eqref{E:BCs_elastoplastic_new} is deduced to Eq. \eqref{E:BCs_elastoplastic}. This is not a rigorous proof since $P_0(p=0^+, V) = 0$ is only proved to be correct when $H \to \infty$. A rigorous proof of $P_0(p=0^+, V) = P_0(p=H^-, V)=0$ becomes much easier using the new solution.
\begin{proof}
Since PDF has non-negative value, $C(V) \geq 0$ according to Eq. \eqref{E:BCs_elastoplastic_new}. Assuming $C(V) > 0$, then we can rewrite Eq. \eqref{E:Nonlinear_elastoplastic_1} with $\bar{p} = H$ as
\begin{equation}\label{E:Relation}
  (1 - a - b)\tilde{P}_0(H, V) = C(V) > 0.
\end{equation}
Given a finite value of $V$, $\tilde{P}_0(H, V) > 0$, then $1 - a - b > 0$. According to the previous asymptotic analysis, as $\bar{p} = H$, $a = 0$ and $b = 1$. This contradicts with $1 - a - b > 0$. Therefore, $C(V)$ must be zero, and the boundary condition, $P_0(p=0^+, V) = P_0(p=H^-, V)=0$.
\end{proof}

The constant hardness assumption is adopted to account for surface plasticity in many elastoplastic rough surface contact models which have been applied to various tribological problems \cite{Almqvist07, Akchurin16, Persson01}. The hardness assumption stems from the work of Tabor \cite{Tabor51} where he found the mean contact pressure on a spherical indenter does not exceed 3$\sigma_Y$. Therefore, the plastic flow on the surface is approximately accounted for in many models by an extra constraint where the contact pressure cannot exceed 3$\sigma_Y$. In the last two decades, the accuracy of the constant hardness model has been challenged by several counter examples \cite{Krithivasan07, Ghaednia17}. Additionally, volume conservation cannot be strictly satisfied under the constant hardness assumption \cite{Weber18}. Recently, Fr\'{e}rot et al. \cite{Frerot20} studied the effect of different types of plasticity model on the prediction of the real contact area and PDF of the contact pressure. Given the same normal load, the more realistic J2 plasticity model with a linear isotropic hardening results in a larger real contact area \cite{Frerot20}. A sharp peak is introduced by the constant hardness assumption in the PDF of the contact pressure, and does not exist for J2 plasticity model. The PDF of the contact pressure whose value is less than the hardness is significantly increased when the constant hardness assumption is replaced by J2 plasticity model. In summary, care should be taken when applying the present Persson theory to elastoplastic rough surface contact. It may underestimate both the real contact area and the PDF of the contact pressure if the material follows J2 plasticity model with strain hardening and the plastic deformation on the interface is dominant \cite{Venugopalan19}. It remains a future goal to find an effective way to extend the Persson’s model to cover elastoplastic contact with strain hardening.

In the above derivation, Persson's theory of elastoplastic contact has only been applied to the contact scheme where a rigid flat is in normal contact with a nominally flat rough surface. For a linear elastic contact, it was proved by Barber \cite{Barber03} that the roughness with surface height $h$ in contact with a rigid flat is equivalent to a roughness with surface height $h_1$ in contact with a roughness with surface height $h_2$ using the plane strain modulus $E^*$ and $h = h_1 + h_2$, as long as root mean square surface gradients $\langle \nabla h_1 \rangle \ll 1$ and $\langle \nabla h_2 \rangle \ll 1$. This equivalent formulation was introduced to Persson's theory of contact by Scaraggi and Persson \cite{Scaraggi15}. For elastoplastic contact, no relevant work has been conducted so far to prove the validity of this equivalent formulation. The closest one is the recent work of Shi and Zou \cite{Shi18}. They found that, (1) if the material model follows J2 plasticity with a linear isotropic hardening, the real area of contact between two rough surfaces predicted by the finite element model is underestimated using the equivalent roughness vs rigid flat scheme (see Fig. 7(a) of \cite{Shi18}) and (2) if the material is elastic-perfectly plastic, both schemes result in the same real area of contact (see Fig. 7(b) of \cite{Shi18}). Since both plasticity models are solid mechanics models, rather than a constant hardness model, it remains unclear whether the equivalent formulation of Scaraggi and Persson can be applied to Persson's theory of elastoplastic contact.

\section{Conclusion}
In this study, Persson's theory of elastoplastic contact with a scale-independent hardness is revisited. The previous solution of $P(p, \xi)$ is in the form of a Fourier series (in terms of an infinite sum of sines). Here we find a simpler form of $P(p, \xi)$ using the superposition of three Gaussian functions. Excellent agreement is found between both solutions. A rigorous proof of the boundary condition is given based on the new solution.

\section*{Declarations}
\subsection*{Funding}
This work was supported by the Leverhulme Trust through Project Grant ``Fundamental mechanical behavior of nano and micro structured interfaces" (RPG-2017-353), the National Natural Science Foundation of China (No. 52105179), the Fundamental Research Funds for the Central Universities of China (No. PA2021KCPY0029) and Jiangsu Key Laboratory of Engineering Mechanics, Southeast University (No. LEM21A03).

\subsection*{Competing Interests}
The authors have no relevant financial or non-financial interests to disclose.

\subsection*{Author Contributions}
{\bf Yang Xu}: Methodology, Conceptualization, Software, Validation, Formal analysis, Visualization, Investigation, Writing – original
draft, Funding acquisition, Supervision, Writing - review and editing. {\bf Xiaobao Li}: Writing - review and editing, Funding acquisition. {\bf Daniel M. Mulvihill}: Methodology, Writing - review and editing, Funding acquisition.

\begin{thebibliography}{99}

\bibitem{Greenwood66}
Greenwood, J.A. and Williamson, J.P., 1966.
Contact of nominally flat surfaces.
Proceedings of the royal society of London. Series A. Mathematical and physical sciences, {\bf 295}(1442), pp.300-319.

\bibitem{Persson01}
Persson, B.N., 2001.
Theory of rubber friction and contact mechanics.
The Journal of Chemical Physics, {\bf 115}(8), pp.3840-3861.

\bibitem{Johnson87}
Johnson, K.L., 1987.
\emph{Contact mechanics}.
Cambridge University Press.

\bibitem{Jackson05}
Jackson, R.L. and Green, I., 2005.
A finite element study of elasto-plastic hemispherical contact against a rigid flat.
ASME Journal of Tribology, {\bf 127}(2), pp.343-354.

\bibitem{Ciavarella06}
Ciavarella, M., Greenwood, J.A. and Paggi, M., 2008.
Inclusion of ``interaction" in the Greenwood and Williamson contact theory.
Wear, {\bf 265}(5-6), pp.729-734.

\bibitem{Afferente12}
Afferrante, L., Carbone, G. and Demelio, G., 2012.
Interacting and coalescing Hertzian asperities: a new multiasperity contact model.
Wear, {\bf 278}, pp.28-33.

\bibitem{Xu14}
Xu, Y., Jackson, R.L. and Marghitu, D.B., 2014.
Statistical model of nearly complete elastic rough surface contact.
International Journal of Solids and Structures, {\bf 51}(5), pp.1075-1088.
Tribology Letters, {\bf 63}(3), p.42.
%
\bibitem{Xu17}
Xu, Y. and Jackson, R.L., 2017.
Statistical models of nearly complete elastic rough surface contact-comparison with numerical solutions.
Tribology International, {\bf 105}, pp.274-291.

\bibitem{Persson01a}
Persson, B.N.J., 2001.
Elastoplastic contact between randomly rough surfaces.
Physical Review Letters, {\bf 87}(11), p.116101.
%
\bibitem{Persson06}
Persson, B.N., 2006.
Contact mechanics for randomly rough surfaces.
Surface Science Reports, {\bf 61}(4), pp.201-227.
%
\bibitem{Venugopalan19}
Venugopalan, S.P., Irani, N. and Nicola, L., 2019.
Plastic contact of self-affine surfaces: Persson’s theory versus discrete dislocation plasticity.
Journal of the Mechanics and Physics of Solids, {\bf 132}, p.103676.
%
\bibitem{Manners06}
Manners, W. and Greenwood, J.A., 2006.
Some observations on Persson's diffusion theory of elastic contact.
Wear, {\bf 261}(5-6), pp.600-610.
%
\bibitem{Wang17}
Wang, A. and M\"{u}ser, M.H., 2017.
Gauging Persson theory on adhesion.
Tribology Letters, {\bf 65}(3), p.103.
%
\bibitem{Persson18}
Persson, B.N.J., 2018.
The dependency of adhesion and friction on electrostatic attraction.
The Journal of Chemical Physics, {\bf 148}(14), p.144701.
%
\bibitem{Persson05}
Persson, B.N., Albohr, O., Tartaglino, U., Volokitin, A.I. and Tosatti, E., 2004.
On the nature of surface roughness with application to contact mechanics, sealing, rubber friction and adhesion.
Journal of physics: Condensed matter, {\bf 17}(1), p.R1.
%
\bibitem{Yastrebov15}
Yastrebov, V.A., Anciaux, G. and Molinari, J.F., 2015.
From infinitesimal to full contact between rough surfaces: evolution of the contact area.
International Journal of Solids and Structures, {\bf 52}, pp.83-102.

\bibitem{Almqvist07}
Almqvist, A., Sahlin, F., Larsson, R., and Glavatskih, S, 2007.
On the dry elasto-plastic contact of nominally flat surfaces.
Tribology International, {\bf 40}(4), pp. 574-579.

\bibitem{Akchurin16}
Akchurin, A., Bosman, R., and Lugt, P. M., 2016.
A stress-criterion-based model for the prediction of the size of wear particles in boundary lubricated contacts.
Tribology Letters, {\bf 64}(3), pp.1-12.

\bibitem{Tabor51}
Tabor, D. 1951. The Hardness of Metals. Clarendon Press, Oxford.

\bibitem{Ghaednia17}
Ghaednia, H., Wang, X., Saha, S., Xu, Y., Sharma, A., and Jackson, R. L., 2017.
A review of elastic–plastic contact mechanics.
Applied Mechanics Reviews, {\bf 69}(6), 060804.

\bibitem{Krithivasan07}
Krithivasan, V., and Jackson, R. L. 2007.
An analysis of three-dimensional elasto-plastic sinusoidal contact.
Tribology Letters, {\bf 27}(1), pp. 31-43.

\bibitem{Weber18}
Weber, B., Suhina, T., Junge, T., Pastewka, L., Brouwer, A. M., and Bonn, D. 2018.
Molecular probes reveal deviations from Amontons’ law in multi-asperity frictional contacts.
Nature Communications, {\bf 9}(1), pp. 1-7.

\bibitem{Frerot20}
Fr\'{e}rot, L., Anciaux, G., and Molinari, J. F. 2020.
Crack nucleation in the adhesive wear of an elastic-plastic half-space.
Journal of the Mechanics and Physics of Solids, {\bf 145}, 104100.

\bibitem{Barber03}
Barber, J.R., 2003.
Bounds on the electrical resistance between contacting elastic rough bodies.
Proceedings of the royal society of London. Series A: mathematical, physical and engineering sciences, {\bf 459}(2029), pp.53-66.
%
\bibitem{Scaraggi15}
Scaraggi, M. and Persson, B.N., 2015.
General contact mechanics theory for randomly rough surfaces with application to rubber friction.
The Journal of Chemical Physics, {\bf 143}(22), p.224111.
%
\bibitem{Shi18}
Shi, X. and Zou, Y., 2018.
A comparative study on equivalent modeling of rough surfaces contact.
ASME Journal of Tribology, {\bf 140}(4), 041402.

\end{thebibliography}
\end{document}